\documentclass[osajnl,twocolumn,showpacs,superscriptaddress,10pt]{revtex4-1}
\usepackage{amsmath,amssymb,graphicx}

\begin{document}

\title{Measuring different types of transverse momentum correlations in the biphoton's Fourier plane}

\author{Omar Calder\'on-Losada}\email{Corresponding author: o.calderon31@uniandes.edu.co}
\author{Jefferson Fl\'orez}
\author{Juan P. Villabona-Monsalve}
\author{Alejandra Valencia}
\affiliation{Laboratorio de \'Optica Cu\'antica, Universidad de los Andes, A.A. 4976, Bogot\'a D.C., Colombia}

\begin{abstract}
In this work, we present a theoretical and experimental study about the spatial correlations of paired photons generated by type-II spontaneous parametric down-conversion. In particular, we show how these correlations can be positive or negative depending on the direction in which the far field plane is scanned and the polarization post-selected. Our results provide a straightforward way to observe different kind of correlations, that complement other well known methods to tune the spatial correlations of paired photons.
\end{abstract}


\maketitle

\section{Introduction}
Correlations have played an important role in fundamental physics and practical applications. Regarding fundamental issues, it was precisely the fact that quantum mechanics allows the existence of the strong correlation, called entanglement, that motivated the discussions between Einstein and Bohr in the first years of quantum physics~\cite{EPR35}. In the 70's , the possibility of measuring correlations between different locations enabled the experimental demonstration of entanglement, revealing in this way new features of nature~\cite{Freedman72}.

Besides the fundamental physical implications behind correlations, they have become a convenient  tool for practical applications. For example, the feature of non-locality derived from entanglement appears as a suitable element for remote applications. In particular, using the continuous time/frequency variables, the possibility of performing remote spectrometry and different protocols for clock synchronization have been developed~\cite{Scarcelli03,Valencia04}. Also, in the spatial case, experiments such as quantum imaging, quantum interference and quantum lithography have been presented~\cite{Strekalov:95,Pittman:95,DAngelo:01,Giovannetti:09}. Furthermore, regarding discrete variables, polarization entanglement has allowed the use of teleportation in quantum information tasks~\cite{Nielsen11}.

Since the 80's, the nonlinear optical process of spontaneous parametric down conversion (SPDC) has provided a convenient source of  entangled photon pairs~\cite{Shih88}. Interestingly, due to the intrinsic nature of the SPDC process the generated photons are entangled in the temporal~\cite{Gisin02}, spatial~\cite{DAngelo04} and polarization degrees of freedom~\cite{Kwiat95}. The models developed so far, to describe the SPDC process, suit very well with the experimental results for the spectral correlations~\cite{Shih2011,Migdall13}. Regarding the spatial domain several authors have reported the dependence of the spatial correlations on crystal's length, pump's waist, geometry of the down-converted process, and even different combinations of the transverse spatial variables~\cite{Ramirez-Alarcon:13,Molina-Terriza05,Procopio14,Osorio:07,Walborn10}. More recently, the implementation of a new family of sensitive CCD cameras has allowed to revisit and study such spatial correlations~\cite{Hamar:10,Edgar:12}, not only in SPDC but also in the high gain regime~\cite{Brambilla:04,Machulka:14}. All these interesting results have been presented for type-I phase matching. On the other hand, for photons generated by non-collinear type-II SPDC there has been less experimental work. For example, in reference~\cite{Ostermeyer09} the position and transverse momentum correlations of photons pairs have been measured in order to study the spatial entanglement. Although in this paper the authors used a type-II crystal, the effects of the different polarizations between signal and idler on the spatial correlations are lost because the setup used combines the SPDC output with a two-photon interferometer. 

In the present work, we report the measurement of the transverse momentum correlation for non-collinear type-II SPDC pairs as they come out directly from the crystal. As a remarkable result, we find different kind of transverse correlations between photons when we scan in perpendicular directions in the Fourier plane and use different post-selection schemes. These experimental results provide a direct way to have different kind of correlations, that complement other well known methods to tune them. Additionally, it could be used to generated a counter-Einstein-Podolsky-Rosen (CEPR) state that has positive correlation in transverse momentum and an negative correlation (anti-correlation) in position~\cite{YunCEPR12}.

\section{Theoretical background}\label{sec:theory}
The physical phenomena studied in the present work are depicted in Fig.~\ref{fig:Expsetup}: photons generated by SPDC are collected by two detection systems placed in the Fourier plane of the crystal's output face in order to measure coincidences between the two detectors and reconstruct the spatial transverse correlation. For a type-II SPDC process, using first order perturbation theory and the paraxial approximation, the two-photon state as a function of the transverse wavevectors $\mathbf{q}_\mu=(q_\mu^x,q_\mu^y)$ and frequency detunings, $\Omega_\mu=\omega_\mu-\omega_0^\mu$, around the central frequencies, $\omega_0^\mu$, for the extraordinary ($e$) and ordinary ($o$) fields ($\mu=e,o$), is given by
\begin{multline}
|\psi\rangle=\int d\mathbf{q}_e d\mathbf{q}_o d\Omega_e d\Omega_o\times\\ \hspace{-15mm} \times \left[\Phi(\mathbf{q}_e,\Omega_e;\mathbf{q}_o,\Omega_o)\hat{a}^\dagger(\Omega_e,\mathbf{q}_e)\hat{a}^\dagger(\Omega_o,\mathbf{q}_o)\right.\\ \left. + \Phi(\mathbf{q}_o,\Omega_o;\mathbf{q}_e,\Omega_e)\hat{a}^\dagger(\Omega_e,\mathbf{q}_e)\hat{a}^\dagger(\Omega_o,\mathbf{q}_o)\right]|0\rangle,
\label{eq:PhotonPairState}
\end{multline}
where $\Phi(\mathbf{q}_e,\Omega_e;\mathbf{q}_o,\Omega_o)$ and $\Phi(\mathbf{q}_o,\Omega_o;\mathbf{q}_e,\Omega_e)$ are the mode functions or biphotons that contain all the information about the correlations between the pair of down-converted photons and the operator $\hat{a}^\dagger$ indicates the creation of a $\mu$-polarized photon with transverse momentum $\mathbf{q}_\mu$ and frequency detuning $\Omega_\mu$.

The mode function $\Phi(\mathbf{q}_e,\Omega_e;\mathbf{q}_o,\Omega_o)$ is related with the joint probability of detecting both an $e$-polarized down-converted photon, with transverse momentum $\mathbf{q}_e$ and frequency detuning $\Omega_e$, at detector $A$ and an $o$-polarized photon, with transverse momentum $\mathbf{q}_o$ and frequency detuning $\Omega_o$ at detector $B$. An analogous definition applies for $\Phi(\mathbf{q}_o,\Omega_o;\mathbf{q}_e,\Omega_e)$. Particularly, $\Phi(\mathbf{q}_e,\Omega_e;\mathbf{q}_o,\Omega_o)$ reads
\begin{multline}
\Phi(\mathbf{q}_e,\Omega_e;\mathbf{q}_o,\Omega_o)=\mathcal{N}\alpha(\Delta_0,\Delta_1)\beta(\Omega_e,\Omega_o)\\
\times\text{sinc}\left(\frac{\Delta_k L}{2}\right)
\exp\left(i\frac{\Delta_k L}{2}\right),
\label{eq:MF}
\end{multline}
where $\mathcal{N}$ is a normalization constant, $\alpha(\Delta_0,\Delta_1)$ has the functional form of the pump's transverse distribution, $L$ is the length of the nonlinear crystal and $\Delta_0$, $\Delta_1$ and $\Delta_k$ are functions that result from the phase matching conditions and are defined as
\begin{subequations}\label{eq:deltas}
\begin{align}
\Delta_0&=q_e^x+q_e^x,\label{delta0}\\
\begin{split}
\Delta_1&=q_e^y\cos\phi_e+q_o^y\cos\phi_o\\
&\qquad{}-N_e\Omega_e\sin\phi_e+N_o\Omega_o\sin\phi_o-\rho_e q_e^x\sin \phi_e,\end{split}\label{delta1}\\
\begin{split}
\Delta_k&=N_p(\Omega_e+\Omega_o)-N_e\Omega_e\cos\phi_e-N_o\Omega_o\cos\phi_o\\
& \qquad -q_e^y\sin \phi_e+q_o^y\sin \phi_o +\rho_p\Delta_0-\rho_e q_e^x\cos \phi_e.\end{split}\label{deltaz}
\end{align}
\end{subequations}

The angles $\phi_e$ and $\phi_o$ are the creation angles of the down-converted photons inside the crystal with respect to the pump's propagation direction, whereas the angles $\rho_p$ and $\rho_e$ account for the walk-off of the pump ($p$) and the extraordinary down-converted photon, respectively. Such angles are given by $\rho_\epsilon=-\frac{1}{n_\epsilon}\frac{\partial n_\epsilon}{\partial \theta_\epsilon}$, ($\epsilon=p,e$), where $n_\epsilon$ is the effective refractive index and $\theta_\epsilon$ is the angle formed by the corresponding wavevector, and the optical axis of the nonlinear crystal. $N_\mu=\frac{d n_\mu(\omega_\mu)\omega_\mu}{d\omega_\mu}\big|_{\omega_\mu=\omega_0^\mu}$ denotes the inverse of the group velocity for each photon.

A good way to grasp the information about the correlations that are contained in Eq.~(\ref{eq:MF}) is by taking into account the following considerations: A pump beam with a Gaussian profile and a waist $\text{w}_p$ in such a way that $\alpha(\Delta_0,\Delta_1)\propto\exp{\left[-\text{w}^2_p(\Delta_0^2+\Delta_1^2)/4\right]}$, and the sinc function approximated by a Gaussian function with the same width at $1/e^2$ of its maximum, i.e. $\text{sinc(x)}\approx\exp{(-\gamma x^2)}$ with $\gamma$ equal $0.193$. With these two considerations Eq.~(\ref{eq:MF}) reduces to
\begin{multline}
\Phi(\mathbf{q}_e,\Omega_e;\mathbf{q}_o,\Omega_o)=\mathcal{N}\times\\
\times\exp{\left[-\frac{\text{w}^2_p(\Delta_0^2+\Delta_1^2)}{4}-\gamma\left(\frac{\Delta_k L}{2}\right)^2+i\frac{\Delta_k L}{2}\right]}.
\label{eq:MFexp}
\end{multline}

In order to observe the transverse correlations, the frequency information has to be traced out, hence the spatial biphoton, $\tilde{\Phi}(\mathbf{q}_e;\mathbf{q}_o)$ is given by $\int d\Omega_e d\Omega_o f_e(\Omega_e)f_o(\Omega_o)\Phi(\mathbf{q}_e,\Omega_e;\mathbf{q}_o,\Omega_o)$, where $f_\mu(\Omega_\mu)$ represents the behavior of the spectral filters placed in front of each detector. For the present study, these filters are modeled as $f_\mu(\Omega_\mu)=\exp\left[-\Omega_\mu^2/(4\sigma^2_\mu)\right]$, with bandwidth $\sigma_\mu$ chosen to achieve a regimen where the spatial-spectral correlations are completely broken~\cite{Florez15}.

Experimentally, it is possible to obtain information about the spatial biphoton by measuring the rate of coincidence counts. For an $e$-photon with transverse momentum $\mathbf{q}_e$ at the detector $A$, and the $o$-photon with transverse momentum $\mathbf{q}_o$ at detector $B$ such rate is given by
\begin{multline}
S(\mathbf{q}_e;\mathbf{q}_o)\equiv|\tilde{\Phi}(\mathbf{q}_e;\mathbf{q}_o)|^2=\\
\left|\int d\Omega_e d\Omega_o f_e(\Omega_e)f_o(\Omega_o)\Phi(\mathbf{q}_e,\Omega_e;\mathbf{q}_o,\Omega_o)\right|^2.
\label{eq:coinrate}
\end{multline}
Analogously, the corresponding coincidence rate where the $e$-photon is detected at $B$ and the $o$-photon is detected at $A$, is defined by $S(\mathbf{q}_o;\mathbf{q}_e)\equiv|\tilde{\Phi}(\mathbf{q}_o,\mathbf{q}_e)|^2$.

The theoretical predictions for the coincidence rates can be seen in the top panels of Figures~\ref{fig:MFHorTeoExp} and~\ref{fig:MFVerTeoExp}, where it has been used the fact that in the Fourier plane there is a one-to-one correspondence between the transverse momentum and position. Explicitly $\mathbf{q}=\frac{2\pi}{\lambda^{0} f}\tilde{\mathbf{x}}$, and therefore
\begin{equation}
S(\mathbf{q}_e;\mathbf{q}_o)=\left|\tilde{\Phi}\left(\frac{2\pi}{\lambda_e^0 f}\tilde{\mathbf{x}}_A,\frac{2\pi}{\lambda_o^0 f}\tilde{\mathbf{x}}_B\right)\right|^2,\label{eq:coin4measure}
\end{equation}
where $\lambda^0_\mu$ is the central wavelength of the $\mu$-polarized down-converted photon, $\tilde{\mathbf{x}}_\eta=(x_\eta,y_\eta)$ ($\eta=A,B$) denotes a position vector in the biphoton's Fourier plane, and $f$ is the focal length of a lens that is used to do the transformation to the Fourier plane in the experimental setup. Fig.~\ref{fig:MFHorTeoExp}(a) and Fig.~\ref{fig:MFHorTeoExp}(b) depict $S(q_e^y;q_o^y)$ and $S(q_o^y;q_e^y)$, respectively, showing the correlations in the $y$-direction between the two Fourier planes. On the other hand, Fig.~\ref{fig:MFVerTeoExp}(a) and Fig.~\ref{fig:MFVerTeoExp}(b) depict $S(q_e^x;q_o^x)$ and $S(q_o^x;q_e^x)$, respectively, showing the correlations in the $x$-direction. It is important to point out three remarkable observations. First, the $y$-transverse momentum exhibits a positive correlation whereas the $x$-transverse momentum shows a negative one. This observation is interesting since it tells us that just by performing a change in the scanning direction of the Fourier plane, it is possible to observe a completely different kind of spatial correlation. In other words, in a type-II SPDC process, both kind of momentum correlations, positive and negative, are present depending on the observed direction of the far-field plane. Second, the positive correlation present in the $y$-direction may be surprising at first sight, since one expects an anti-correlation for the transverse momentum of down-converted photons; however, this is true only for certain values of the pump's waist. The waist chosen to depict Fig.~\ref{fig:MFHorTeoExp}, $\text{w}_p=31$ $\mu$m, is such that together with the other relevant parameters in the SPDC generation yields to a positive correlation. This is similar to what has been reported for the type-I case~\cite{Molina-Terriza05}. Third, the spatial correlation in the $y$-direction is insensitive to which polarization arrive to each detector. On the contrary, the spatial correlation in the $x$-direction is highly affected by the detected polarization as is revealed by the different orientations of $S(q_e^x;q_o^x)$ and $S(q_o^x;q_e^x)$.

\section{Experimental realization}\label{sec:experiment}
\begin{figure}
\centering
\fbox{\includegraphics[width=0.45\textwidth]{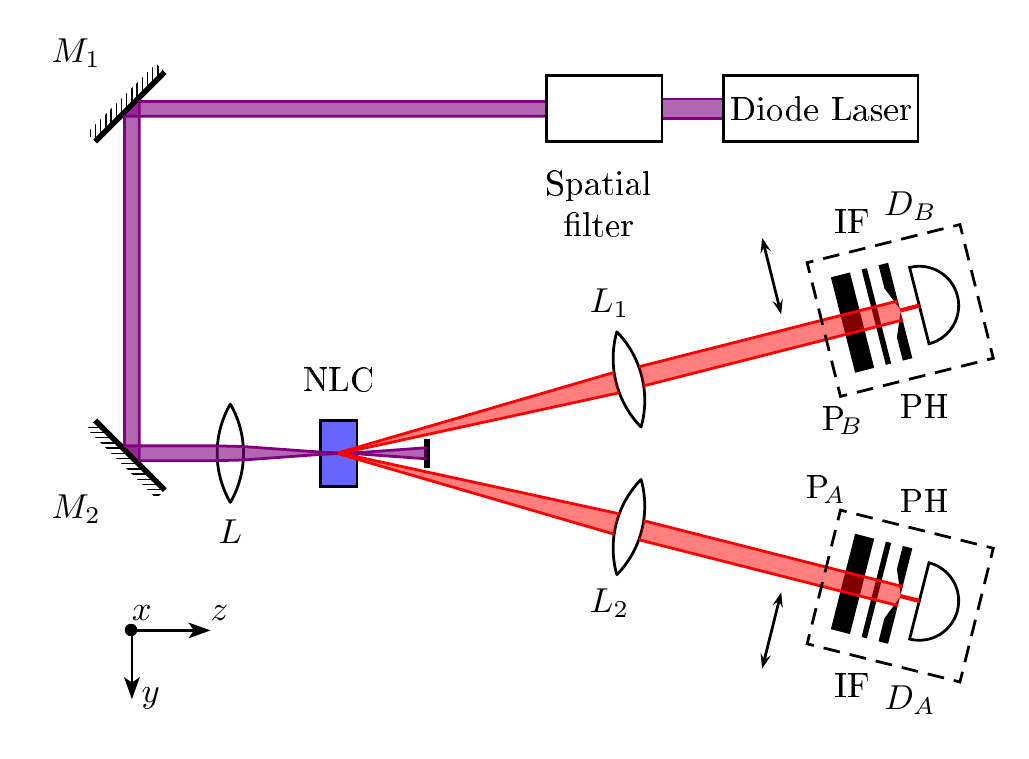}}
\caption{Experimental setup. The diode laser has central wavelength at 407 nm and is spatially shaped as a Gaussian beam with a waist of 31 $\mu$m and 42 $\mu$m in the horizontal ($y$-direction) and vertical ($x$-direction) direction, respectively. $M_1$ and $M_2$ are mirrors. $L$ is a lens with focal length of 150 mm that fixes the desired waist on the 4-mm BBO type-II crystal (NLC) to generate a pair of down-converted photons at a half open angle of $\sim6^\circ$. Each lens $L_1$ and $L_2$, both with focal lengths of 750 mm, configures a 2$f$-system. The detection system consists of a polarizer $\text{P}_\eta$, a 5 nm-bandwidth interference filter (IF) centered at 814 nm, a pinhole (PH) and a multimode fiber to take the photons to detectors, $D_A$ and $D_B$.}
\label{fig:Expsetup}
\end{figure}

To corroborate the previous theory, the setup in Fig.~\ref{fig:Expsetup} is implemented, where a diode laser beam, centered at 407 nm and spatially shaped by a spatial filter, is focused by a lens $L$ into a 4-mm-length type-II BBO crystal. Each of the generated down-converted photons passes through a 2$f$-system in order to obtain the momentum distribution of the photon pairs at the Fourier plane. The combined scan to measure momentum correlations is accomplished by two detection systems that consist on polarizers, an interference filters, 2.0 mm-diameter pinholes, multimode fibers and single photon counters (SPCM-AQRH-13), $D_A$ and $D_B$, whose outputs are analyzed by a FPGA card that allows to count singles and coincidences. The detection systems are mounted on automated translational stages to do a full scan of the biphoton's Fourier plane either in the horizontal or vertical direction.
In the experiment, we perform the measurement of $S(\mathbf{q}_e;\mathbf{q}_o)$ or $S(\mathbf{q}_o;\mathbf{q}_e)$ by rotating the polarizers, $\text{P}_A$ and $\text{P}_B$, to detect either the $e$- or $o$- photon in the corresponding detector, $D_A$ or $D_B$.
\begin{figure}
\centering
\fbox{\includegraphics[width=0.45\textwidth]{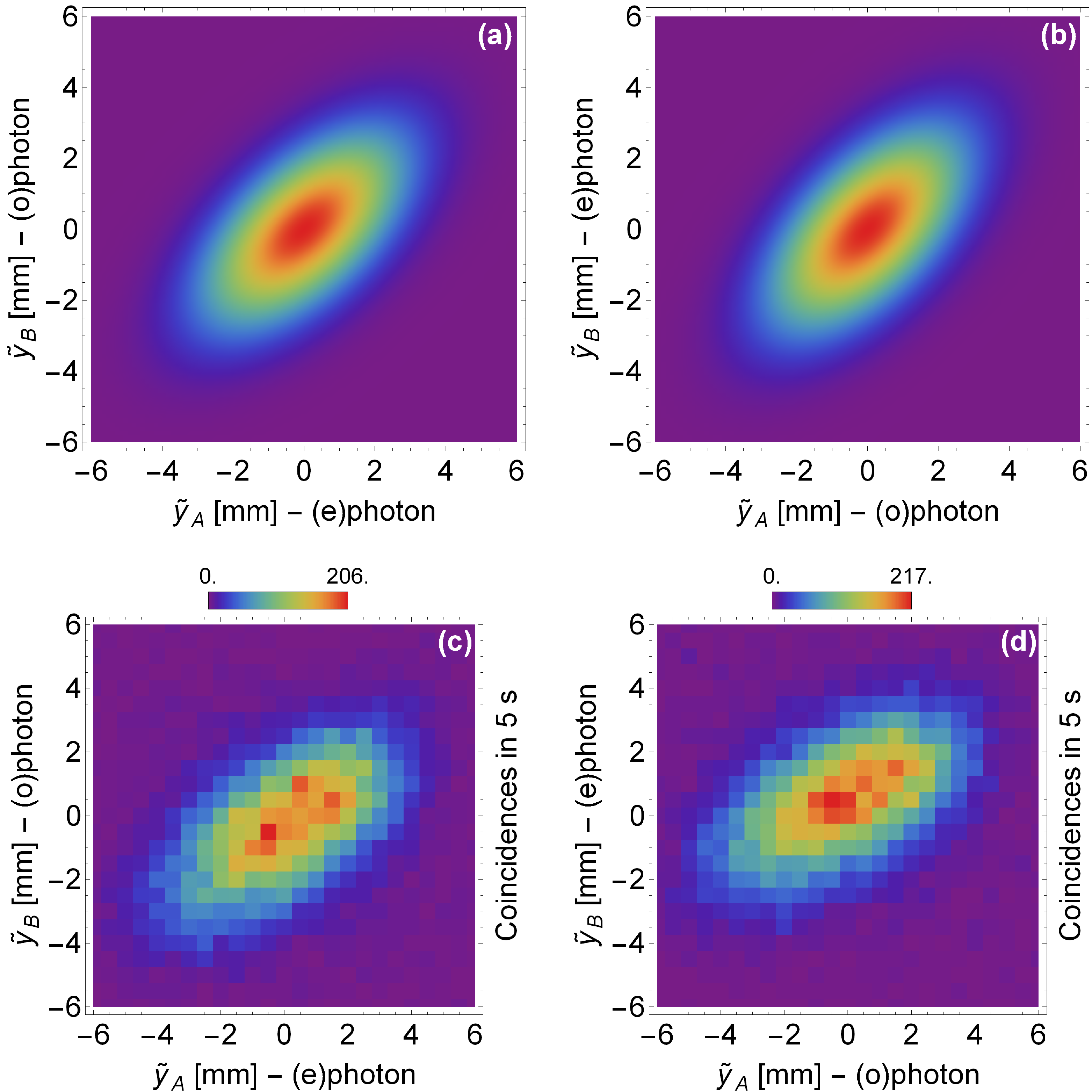}}
\caption{Spatial correlations in the $y$-direction at the biphoton's Fourier plane. Panels (a) and (c) depict $S(q_e^y;q_o^y)$ from the theoretical model and the experimental results, respectively. Panels (b) and (d) depict $S(q_o^y;q_e^y)$ from the theoretical model and the experimental results, respectively.}
\label{fig:MFHorTeoExp}
\end{figure}

\begin{figure}
\centering
\fbox{\includegraphics[width=0.45\textwidth]{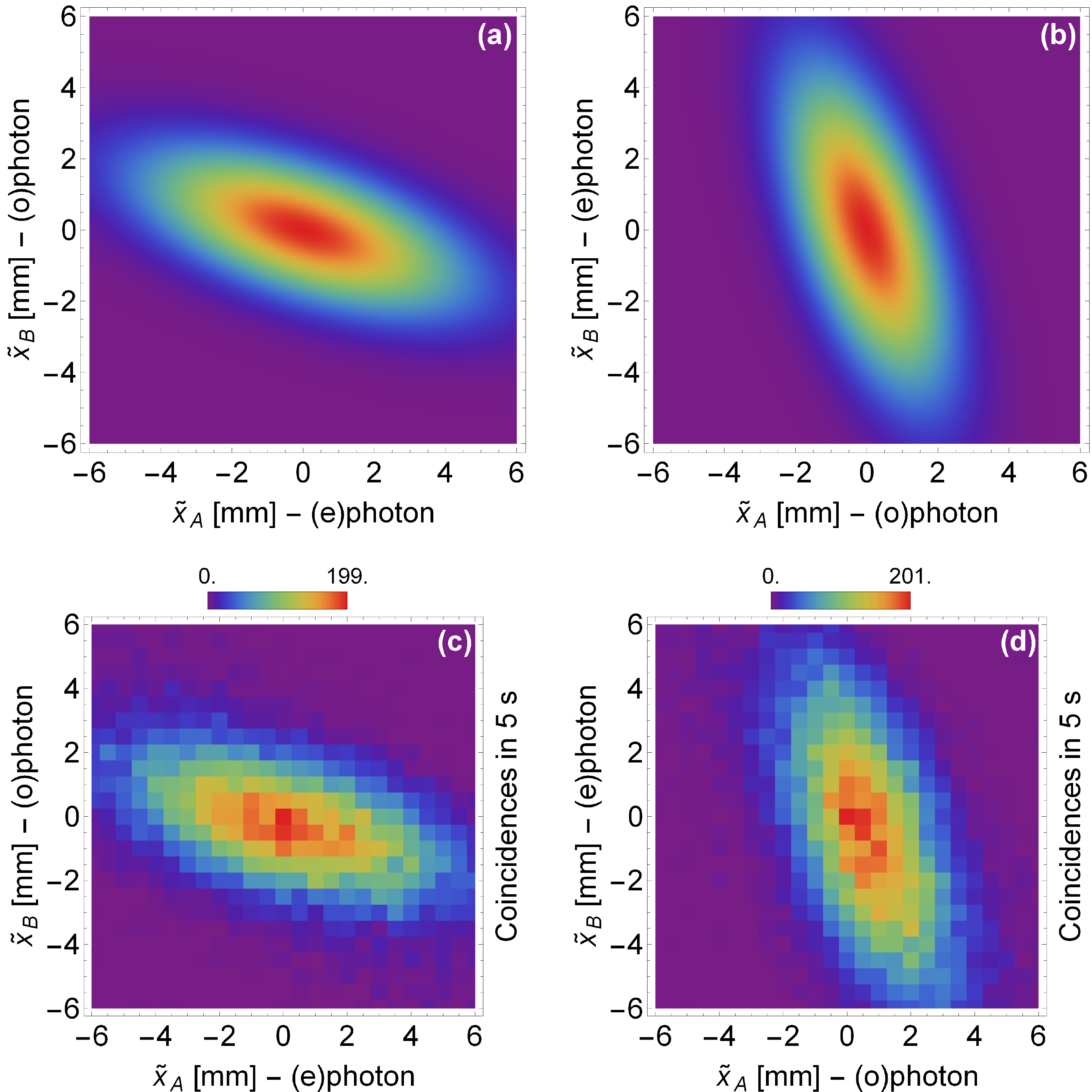}}
\caption{Spatial correlations in the $x$-direction at the biphoton's Fourier plane. Panels (a) and (c) depict $S(q_e^x;q_o^x)$ from the theoretical model and the experimental results, respectively. Panels (b) and (d) depict $S(q_o^x;q_e^x)$ from the theoretical model and the experimental results, respectively.}
\label{fig:MFVerTeoExp}
\end{figure}

When both detectors are scanned in the direction parallel to the optical table ($y$-direction), the bottom panel in Fig.~\ref{fig:MFHorTeoExp} shows the experimental results post-selecting different polarizations for each detector. In particular, Fig.~\ref{fig:MFHorTeoExp}(c) shows $S(q_e^y;q_o^y)$ and Fig.~\ref{fig:MFHorTeoExp}(d) shows $S(q_o^y,q_e^y)$. In both situations, the momentum correlation is in good agreement with the theory. On the other hand, the bottom panels of Fig.~\ref{fig:MFVerTeoExp} display the transverse momentum correlations when the detectors scan the Fourier plane in the  perpendicular direction to the optical table ($x$-direction) for different post-selected polarizations. Fig.~\ref{fig:MFVerTeoExp}(c) depicts $S(q_e^x;q_o^x)$, whereas Fig.~\ref{fig:MFVerTeoExp}(d) illustrates $S(q_o^x;q_e^x)$. From the graphs, it is clear that the vertical transverse correlation changes drastically with respect to the  horizontal transverse behavior: The orientation of the coincidence rate shows a negative correlation, mainly due to the contribution of the walk-off angles, $\rho_{\mu}$, only in the $x$-direction as can be seen by the term $\rho_e q_e^x\sin \phi_e$ in Eq.~\ref{eq:deltas}(b) and the term $\rho_p\Delta_0-\rho_e q_e^x\cos \phi_e$ in Eq.~\ref{eq:deltas}(c). Additionally, it is clear, that the polarizers play an important role in the shape of the spatial mode function, as expected from the theory.

In conclusion, we have experimentally demonstrated that for a fixed value of the pump's waist, it is possible to achieve different kind of transverse correlations between type-II SPDC photons by changing the direction in which the Fourier Plane is scanned. Additionally, the correlation in the vertical direction is affected by the chosen post-selection polarization scheme. These behaviors occur because of the crystal birefringence that results in a walk-off angle that affects differently the $e$- and $o$- photon and introduces a distinguishability between different directions of the transverse correlations. Our results complement the already well known methods in which the pump spatial and spectral profiles are used to tune the spatial correlations of SPDC photons.

\section*{Funding Information and Acknowledgments}
This work was supported and founded by Facultad de Ciencias and Vicerector\'ia de Investigaciones of Universidad de los Andes, Bogot\'a, Colombia.
The authors thank the discussions, contributions and electronic technical support given by David A. Guzm\'an.

\bibliography{biblio_experi-controlcorrelation}

\begin{thebibliography}{27}%
\makeatletter
\providecommand \@ifxundefined [1]{%
 \@ifx{#1\undefined}
}%
\providecommand \@ifnum [1]{%
 \ifnum #1\expandafter \@firstoftwo
 \else \expandafter \@secondoftwo
 \fi
}%
\providecommand \@ifx [1]{%
 \ifx #1\expandafter \@firstoftwo
 \else \expandafter \@secondoftwo
 \fi
}%
\providecommand \natexlab [1]{#1}%
\providecommand \enquote  [1]{``#1''}%
\providecommand \bibnamefont  [1]{#1}%
\providecommand \bibfnamefont [1]{#1}%
\providecommand \citenamefont [1]{#1}%
\providecommand \href@noop [0]{\@secondoftwo}%
\providecommand \href [0]{\begingroup \@sanitize@url \@href}%
\providecommand \@href[1]{\@@startlink{#1}\@@href}%
\providecommand \@@href[1]{\endgroup#1\@@endlink}%
\providecommand \@sanitize@url [0]{\catcode `\\12\catcode `\$12\catcode
  `\&12\catcode `\#12\catcode `\^12\catcode `\_12\catcode `\%12\relax}%
\providecommand \@@startlink[1]{}%
\providecommand \@@endlink[0]{}%
\providecommand \url  [0]{\begingroup\@sanitize@url \@url }%
\providecommand \@url [1]{\endgroup\@href {#1}{\urlprefix }}%
\providecommand \urlprefix  [0]{URL }%
\providecommand \Eprint [0]{\href }%
\providecommand \doibase [0]{http://dx.doi.org/}%
\providecommand \selectlanguage [0]{\@gobble}%
\providecommand \bibinfo  [0]{\@secondoftwo}%
\providecommand \bibfield  [0]{\@secondoftwo}%
\providecommand \translation [1]{[#1]}%
\providecommand \BibitemOpen [0]{}%
\providecommand \bibitemStop [0]{}%
\providecommand \bibitemNoStop [0]{.\EOS\space}%
\providecommand \EOS [0]{\spacefactor3000\relax}%
\providecommand \BibitemShut  [1]{\csname bibitem#1\endcsname}%
\let\auto@bib@innerbib\@empty
\bibitem [{\citenamefont {Einstein}\ \emph {et~al.}(1935)\citenamefont
  {Einstein}, \citenamefont {Podolsky},\ and\ \citenamefont {Rosen}}]{EPR35}%
  \BibitemOpen
  \bibfield  {author} {\bibinfo {author} {\bibfnamefont {A.}~\bibnamefont
  {Einstein}}, \bibinfo {author} {\bibfnamefont {B.}~\bibnamefont {Podolsky}},
  \ and\ \bibinfo {author} {\bibfnamefont {N.}~\bibnamefont {Rosen}},\ }\href
  {\doibase 10.1103/PhysRev.47.777} {\bibfield  {journal} {\bibinfo  {journal}
  {Phys. Rev.}\ }\textbf {\bibinfo {volume} {47}},\ \bibinfo {pages} {777}
  (\bibinfo {year} {1935})}\BibitemShut {NoStop}%
\bibitem [{\citenamefont {Freedman}\ and\ \citenamefont
  {Clauser}(1972)}]{Freedman72}%
  \BibitemOpen
  \bibfield  {author} {\bibinfo {author} {\bibfnamefont {S.~J.}\ \bibnamefont
  {Freedman}}\ and\ \bibinfo {author} {\bibfnamefont {J.~F.}\ \bibnamefont
  {Clauser}},\ }\href {\doibase 10.1103/PhysRevLett.28.938} {\bibfield
  {journal} {\bibinfo  {journal} {Phys. Rev. Lett.}\ }\textbf {\bibinfo
  {volume} {28}},\ \bibinfo {pages} {938} (\bibinfo {year} {1972})}\BibitemShut
  {NoStop}%
\bibitem [{\citenamefont {Scarcelli}\ \emph {et~al.}(2003)\citenamefont
  {Scarcelli}, \citenamefont {Valencia}, \citenamefont {Gompers},\ and\
  \citenamefont {Shih}}]{Scarcelli03}%
  \BibitemOpen
  \bibfield  {author} {\bibinfo {author} {\bibfnamefont {G.}~\bibnamefont
  {Scarcelli}}, \bibinfo {author} {\bibfnamefont {A.}~\bibnamefont {Valencia}},
  \bibinfo {author} {\bibfnamefont {S.}~\bibnamefont {Gompers}}, \ and\
  \bibinfo {author} {\bibfnamefont {Y.}~\bibnamefont {Shih}},\ }\href {\doibase
  http://dx.doi.org/10.1063/1.1637131} {\bibfield  {journal} {\bibinfo
  {journal} {Applied Physics Letters}\ }\textbf {\bibinfo {volume} {83}},\
  \bibinfo {pages} {5560} (\bibinfo {year} {2003})}\BibitemShut {NoStop}%
\bibitem [{\citenamefont {Valencia}\ \emph {et~al.}(2004)\citenamefont
  {Valencia}, \citenamefont {Scarcelli},\ and\ \citenamefont
  {Shih}}]{Valencia04}%
  \BibitemOpen
  \bibfield  {author} {\bibinfo {author} {\bibfnamefont {A.}~\bibnamefont
  {Valencia}}, \bibinfo {author} {\bibfnamefont {G.}~\bibnamefont {Scarcelli}},
  \ and\ \bibinfo {author} {\bibfnamefont {Y.}~\bibnamefont {Shih}},\ }\href
  {\doibase http://dx.doi.org/10.1063/1.1797561} {\bibfield  {journal}
  {\bibinfo  {journal} {Applied Physics Letters}\ }\textbf {\bibinfo {volume}
  {85}},\ \bibinfo {pages} {2655} (\bibinfo {year} {2004})}\BibitemShut
  {NoStop}%
\bibitem [{\citenamefont {Strekalov}\ \emph {et~al.}(1995)\citenamefont
  {Strekalov}, \citenamefont {Sergienko}, \citenamefont {Klyshko},\ and\
  \citenamefont {Shih}}]{Strekalov:95}%
  \BibitemOpen
  \bibfield  {author} {\bibinfo {author} {\bibfnamefont {D.~V.}\ \bibnamefont
  {Strekalov}}, \bibinfo {author} {\bibfnamefont {A.~V.}\ \bibnamefont
  {Sergienko}}, \bibinfo {author} {\bibfnamefont {D.~N.}\ \bibnamefont
  {Klyshko}}, \ and\ \bibinfo {author} {\bibfnamefont {Y.~H.}\ \bibnamefont
  {Shih}},\ }\href {\doibase 10.1103/PhysRevLett.74.3600} {\bibfield  {journal}
  {\bibinfo  {journal} {Phys. Rev. Lett.}\ }\textbf {\bibinfo {volume} {74}},\
  \bibinfo {pages} {3600} (\bibinfo {year} {1995})}\BibitemShut {NoStop}%
\bibitem [{\citenamefont {Pittman}\ \emph {et~al.}(1995)\citenamefont
  {Pittman}, \citenamefont {Shih}, \citenamefont {Strekalov},\ and\
  \citenamefont {Sergienko}}]{Pittman:95}%
  \BibitemOpen
  \bibfield  {author} {\bibinfo {author} {\bibfnamefont {T.~B.}\ \bibnamefont
  {Pittman}}, \bibinfo {author} {\bibfnamefont {Y.~H.}\ \bibnamefont {Shih}},
  \bibinfo {author} {\bibfnamefont {D.~V.}\ \bibnamefont {Strekalov}}, \ and\
  \bibinfo {author} {\bibfnamefont {A.~V.}\ \bibnamefont {Sergienko}},\ }\href
  {\doibase 10.1103/PhysRevA.52.R3429} {\bibfield  {journal} {\bibinfo
  {journal} {Phys. Rev. A}\ }\textbf {\bibinfo {volume} {52}},\ \bibinfo
  {pages} {R3429} (\bibinfo {year} {1995})}\BibitemShut {NoStop}%
\bibitem [{\citenamefont {D'Angelo}\ \emph {et~al.}(2001)\citenamefont
  {D'Angelo}, \citenamefont {Chekhova},\ and\ \citenamefont
  {Shih}}]{DAngelo:01}%
  \BibitemOpen
  \bibfield  {author} {\bibinfo {author} {\bibfnamefont {M.}~\bibnamefont
  {D'Angelo}}, \bibinfo {author} {\bibfnamefont {M.~V.}\ \bibnamefont
  {Chekhova}}, \ and\ \bibinfo {author} {\bibfnamefont {Y.}~\bibnamefont
  {Shih}},\ }\href {\doibase 10.1103/PhysRevLett.87.013602} {\bibfield
  {journal} {\bibinfo  {journal} {Phys. Rev. Lett.}\ }\textbf {\bibinfo
  {volume} {87}},\ \bibinfo {pages} {013602} (\bibinfo {year}
  {2001})}\BibitemShut {NoStop}%
\bibitem [{\citenamefont {Giovannetti}\ \emph {et~al.}(2009)\citenamefont
  {Giovannetti}, \citenamefont {Lloyd}, \citenamefont {Maccone},\ and\
  \citenamefont {Shapiro}}]{Giovannetti:09}%
  \BibitemOpen
  \bibfield  {author} {\bibinfo {author} {\bibfnamefont {V.}~\bibnamefont
  {Giovannetti}}, \bibinfo {author} {\bibfnamefont {S.}~\bibnamefont {Lloyd}},
  \bibinfo {author} {\bibfnamefont {L.}~\bibnamefont {Maccone}}, \ and\
  \bibinfo {author} {\bibfnamefont {J.~H.}\ \bibnamefont {Shapiro}},\ }\href
  {\doibase 10.1103/PhysRevA.79.013827} {\bibfield  {journal} {\bibinfo
  {journal} {Phys. Rev. A}\ }\textbf {\bibinfo {volume} {79}},\ \bibinfo
  {pages} {013827} (\bibinfo {year} {2009})}\BibitemShut {NoStop}%
\bibitem [{Nie(2011)}]{Nielsen11}%
  \BibitemOpen
  \href@noop {} {\emph {\bibinfo {title} {Quantum Computation and Quantum
  Information}}}\ (\bibinfo  {publisher} {Cambridge University Press},\
  \bibinfo {year} {2011})\BibitemShut {NoStop}%
\bibitem [{\citenamefont {Shih}\ and\ \citenamefont {Alley}(1988)}]{Shih88}%
  \BibitemOpen
  \bibfield  {author} {\bibinfo {author} {\bibfnamefont {Y.~H.}\ \bibnamefont
  {Shih}}\ and\ \bibinfo {author} {\bibfnamefont {C.~O.}\ \bibnamefont
  {Alley}},\ }\href {\doibase 10.1103/PhysRevLett.61.2921} {\bibfield
  {journal} {\bibinfo  {journal} {Phys. Rev. Lett.}\ }\textbf {\bibinfo
  {volume} {61}},\ \bibinfo {pages} {2921} (\bibinfo {year}
  {1988})}\BibitemShut {NoStop}%
\bibitem [{\citenamefont {Marcikic}\ \emph {et~al.}(2002)\citenamefont
  {Marcikic}, \citenamefont {de~Riedmatten}, \citenamefont {Tittel},
  \citenamefont {Scarani}, \citenamefont {Zbinden},\ and\ \citenamefont
  {Gisin}}]{Gisin02}%
  \BibitemOpen
  \bibfield  {author} {\bibinfo {author} {\bibfnamefont {I.}~\bibnamefont
  {Marcikic}}, \bibinfo {author} {\bibfnamefont {H.}~\bibnamefont
  {de~Riedmatten}}, \bibinfo {author} {\bibfnamefont {W.}~\bibnamefont
  {Tittel}}, \bibinfo {author} {\bibfnamefont {V.}~\bibnamefont {Scarani}},
  \bibinfo {author} {\bibfnamefont {H.}~\bibnamefont {Zbinden}}, \ and\
  \bibinfo {author} {\bibfnamefont {N.}~\bibnamefont {Gisin}},\ }\href
  {\doibase 10.1103/PhysRevA.66.062308} {\bibfield  {journal} {\bibinfo
  {journal} {Phys. Rev. A}\ }\textbf {\bibinfo {volume} {66}},\ \bibinfo
  {pages} {062308} (\bibinfo {year} {2002})}\BibitemShut {NoStop}%
\bibitem [{\citenamefont {D'Angelo}\ \emph {et~al.}(2004)\citenamefont
  {D'Angelo}, \citenamefont {Kim}, \citenamefont {Kulik},\ and\ \citenamefont
  {Shih}}]{DAngelo04}%
  \BibitemOpen
  \bibfield  {author} {\bibinfo {author} {\bibfnamefont {M.}~\bibnamefont
  {D'Angelo}}, \bibinfo {author} {\bibfnamefont {Y.-H.}\ \bibnamefont {Kim}},
  \bibinfo {author} {\bibfnamefont {S.~P.}\ \bibnamefont {Kulik}}, \ and\
  \bibinfo {author} {\bibfnamefont {Y.}~\bibnamefont {Shih}},\ }\href {\doibase
  10.1103/PhysRevLett.92.233601} {\bibfield  {journal} {\bibinfo  {journal}
  {Phys. Rev. Lett.}\ }\textbf {\bibinfo {volume} {92}},\ \bibinfo {pages}
  {233601} (\bibinfo {year} {2004})}\BibitemShut {NoStop}%
\bibitem [{\citenamefont {Kwiat}\ \emph {et~al.}(1995)\citenamefont {Kwiat},
  \citenamefont {Mattle}, \citenamefont {Weinfurter}, \citenamefont
  {Zeilinger}, \citenamefont {Sergienko},\ and\ \citenamefont
  {Shih}}]{Kwiat95}%
  \BibitemOpen
  \bibfield  {author} {\bibinfo {author} {\bibfnamefont {P.~G.}\ \bibnamefont
  {Kwiat}}, \bibinfo {author} {\bibfnamefont {K.}~\bibnamefont {Mattle}},
  \bibinfo {author} {\bibfnamefont {H.}~\bibnamefont {Weinfurter}}, \bibinfo
  {author} {\bibfnamefont {A.}~\bibnamefont {Zeilinger}}, \bibinfo {author}
  {\bibfnamefont {A.~V.}\ \bibnamefont {Sergienko}}, \ and\ \bibinfo {author}
  {\bibfnamefont {Y.}~\bibnamefont {Shih}},\ }\href {\doibase
  10.1103/PhysRevLett.75.4337} {\bibfield  {journal} {\bibinfo  {journal}
  {Phys. Rev. Lett.}\ }\textbf {\bibinfo {volume} {75}},\ \bibinfo {pages}
  {4337} (\bibinfo {year} {1995})}\BibitemShut {NoStop}%
\bibitem [{Shi(2011)}]{Shih2011}%
  \BibitemOpen
  \href@noop {} {\emph {\bibinfo {title} {An Introduction to Quantum Optics:
  Photon and Biphoton Physics}}}\ (\bibinfo  {publisher} {CRC press},\ \bibinfo
  {year} {2011})\BibitemShut {NoStop}%
\bibitem [{Mig(2013)}]{Migdall13}%
  \BibitemOpen
  \href@noop {} {\emph {\bibinfo {title} {Single-Photon Generation and
  Detection}}}\ (\bibinfo  {publisher} {Academic press},\ \bibinfo {year}
  {2013})\BibitemShut {NoStop}%
\bibitem [{\citenamefont {Ramírez-Alarcón}\ \emph {et~al.}(2013)\citenamefont
  {Ramírez-Alarcón}, \citenamefont {Cruz-Ramírez},\ and\ \citenamefont
  {U’Ren}}]{Ramirez-Alarcon:13}%
  \BibitemOpen
  \bibfield  {author} {\bibinfo {author} {\bibfnamefont {R.}~\bibnamefont
  {Ramírez-Alarcón}}, \bibinfo {author} {\bibfnamefont {H.}~\bibnamefont
  {Cruz-Ramírez}}, \ and\ \bibinfo {author} {\bibfnamefont {A.~B.}\
  \bibnamefont {U’Ren}},\ }\href
  {http://stacks.iop.org/1555-6611/23/i=5/a=055204} {\bibfield  {journal}
  {\bibinfo  {journal} {Laser Physics}\ }\textbf {\bibinfo {volume} {23}},\
  \bibinfo {pages} {055204} (\bibinfo {year} {2013})}\BibitemShut {NoStop}%
\bibitem [{\citenamefont {Molina-Terriza}\ \emph {et~al.}(2005)\citenamefont
  {Molina-Terriza}, \citenamefont {Minardi}, \citenamefont {Deyanova},
  \citenamefont {Osorio}, \citenamefont {Hendrych},\ and\ \citenamefont
  {Torres}}]{Molina-Terriza05}%
  \BibitemOpen
  \bibfield  {author} {\bibinfo {author} {\bibfnamefont {G.}~\bibnamefont
  {Molina-Terriza}}, \bibinfo {author} {\bibfnamefont {S.}~\bibnamefont
  {Minardi}}, \bibinfo {author} {\bibfnamefont {Y.}~\bibnamefont {Deyanova}},
  \bibinfo {author} {\bibfnamefont {C.~I.}\ \bibnamefont {Osorio}}, \bibinfo
  {author} {\bibfnamefont {M.}~\bibnamefont {Hendrych}}, \ and\ \bibinfo
  {author} {\bibfnamefont {J.~P.}\ \bibnamefont {Torres}},\ }\href {\doibase
  10.1103/PhysRevA.72.065802} {\bibfield  {journal} {\bibinfo  {journal} {Phys.
  Rev. A}\ }\textbf {\bibinfo {volume} {72}},\ \bibinfo {pages} {065802}
  (\bibinfo {year} {2005})}\BibitemShut {NoStop}%
\bibitem [{\citenamefont {Procopio}\ \emph {et~al.}(2015)\citenamefont
  {Procopio}, \citenamefont {Rosas-Ortiz},\ and\ \citenamefont
  {Velázquez}}]{Procopio14}%
  \BibitemOpen
  \bibfield  {author} {\bibinfo {author} {\bibfnamefont {L.~M.}\ \bibnamefont
  {Procopio}}, \bibinfo {author} {\bibfnamefont {O.}~\bibnamefont
  {Rosas-Ortiz}}, \ and\ \bibinfo {author} {\bibfnamefont {V.}~\bibnamefont
  {Velázquez}},\ }\href {\doibase 10.1002/mma.3192} {\bibfield  {journal}
  {\bibinfo  {journal} {Mathematical Methods in the Applied Sciences}\ }\textbf
  {\bibinfo {volume} {38}},\ \bibinfo {pages} {2053} (\bibinfo {year}
  {2015})}\BibitemShut {NoStop}%
\bibitem [{\citenamefont {Osorio}\ \emph {et~al.}(2007)\citenamefont {Osorio},
  \citenamefont {Molina-Terriza}, \citenamefont {Font},\ and\ \citenamefont
  {Torres}}]{Osorio:07}%
  \BibitemOpen
  \bibfield  {author} {\bibinfo {author} {\bibfnamefont {C.~I.}\ \bibnamefont
  {Osorio}}, \bibinfo {author} {\bibfnamefont {G.}~\bibnamefont
  {Molina-Terriza}}, \bibinfo {author} {\bibfnamefont {B.~G.}\ \bibnamefont
  {Font}}, \ and\ \bibinfo {author} {\bibfnamefont {J.~P.}\ \bibnamefont
  {Torres}},\ }\href {\doibase 10.1364/OE.15.014636} {\bibfield  {journal}
  {\bibinfo  {journal} {Opt. Express}\ }\textbf {\bibinfo {volume} {15}},\
  \bibinfo {pages} {14636} (\bibinfo {year} {2007})}\BibitemShut {NoStop}%
\bibitem [{\citenamefont {Walborn}\ \emph {et~al.}(2010)\citenamefont
  {Walborn}, \citenamefont {Monken}, \citenamefont {Pádua},\ and\
  \citenamefont {Ribeiro}}]{Walborn10}%
  \BibitemOpen
  \bibfield  {author} {\bibinfo {author} {\bibfnamefont {S.}~\bibnamefont
  {Walborn}}, \bibinfo {author} {\bibfnamefont {C.}~\bibnamefont {Monken}},
  \bibinfo {author} {\bibfnamefont {S.}~\bibnamefont {Pádua}}, \ and\ \bibinfo
  {author} {\bibfnamefont {P.~S.}\ \bibnamefont {Ribeiro}},\ }\href {\doibase
  http://dx.doi.org/10.1016/j.physrep.2010.06.003} {\bibfield  {journal}
  {\bibinfo  {journal} {Physics Reports}\ }\textbf {\bibinfo {volume} {495}},\
  \bibinfo {pages} {87 } (\bibinfo {year} {2010})}\BibitemShut {NoStop}%
\bibitem [{\citenamefont {Hamar}\ \emph {et~al.}(2010)\citenamefont {Hamar},
  \citenamefont {Pe\ifmmode~\check{r}\else \v{r}\fi{}ina}, \citenamefont
  {Haderka},\ and\ \citenamefont {Mich\'alek}}]{Hamar:10}%
  \BibitemOpen
  \bibfield  {author} {\bibinfo {author} {\bibfnamefont {M.}~\bibnamefont
  {Hamar}}, \bibinfo {author} {\bibfnamefont {J.}~\bibnamefont
  {Pe\ifmmode~\check{r}\else \v{r}\fi{}ina}}, \bibinfo {author} {\bibfnamefont
  {O.~c.~v.}\ \bibnamefont {Haderka}}, \ and\ \bibinfo {author} {\bibfnamefont
  {V.}~\bibnamefont {Mich\'alek}},\ }\href {\doibase
  10.1103/PhysRevA.81.043827} {\bibfield  {journal} {\bibinfo  {journal} {Phys.
  Rev. A}\ }\textbf {\bibinfo {volume} {81}},\ \bibinfo {pages} {043827}
  (\bibinfo {year} {2010})}\BibitemShut {NoStop}%
\bibitem [{\citenamefont {Edgar}\ \emph {et~al.}(2012)\citenamefont {Edgar},
  \citenamefont {Tasca}, \citenamefont {Izdebski}, \citenamefont {Warburton},
  \citenamefont {Leach}, \citenamefont {Agnew}, \citenamefont {Buller},
  \citenamefont {Boyd},\ and\ \citenamefont {Padgett}}]{Edgar:12}%
  \BibitemOpen
  \bibfield  {author} {\bibinfo {author} {\bibfnamefont {M.~P.}\ \bibnamefont
  {Edgar}}, \bibinfo {author} {\bibfnamefont {D.~S.}\ \bibnamefont {Tasca}},
  \bibinfo {author} {\bibfnamefont {F.}~\bibnamefont {Izdebski}}, \bibinfo
  {author} {\bibfnamefont {R.~E.}\ \bibnamefont {Warburton}}, \bibinfo {author}
  {\bibfnamefont {J.}~\bibnamefont {Leach}}, \bibinfo {author} {\bibfnamefont
  {M.}~\bibnamefont {Agnew}}, \bibinfo {author} {\bibfnamefont {G.~S.}\
  \bibnamefont {Buller}}, \bibinfo {author} {\bibfnamefont {R.~W.}\
  \bibnamefont {Boyd}}, \ and\ \bibinfo {author} {\bibfnamefont {M.~J.}\
  \bibnamefont {Padgett}},\ }\href {\doibase 10.1038/ncomms1988} {\bibfield
  {journal} {\bibinfo  {journal} {Nat Commun}\ }\textbf {\bibinfo {volume}
  {3}},\ \bibinfo {pages} {984} (\bibinfo {year} {2012})}\BibitemShut {NoStop}%
\bibitem [{\citenamefont {Brambilla}\ \emph {et~al.}(2004)\citenamefont
  {Brambilla}, \citenamefont {Gatti}, \citenamefont {Bache},\ and\
  \citenamefont {Lugiato}}]{Brambilla:04}%
  \BibitemOpen
  \bibfield  {author} {\bibinfo {author} {\bibfnamefont {E.}~\bibnamefont
  {Brambilla}}, \bibinfo {author} {\bibfnamefont {A.}~\bibnamefont {Gatti}},
  \bibinfo {author} {\bibfnamefont {M.}~\bibnamefont {Bache}}, \ and\ \bibinfo
  {author} {\bibfnamefont {L.~A.}\ \bibnamefont {Lugiato}},\ }\href {\doibase
  10.1103/PhysRevA.69.023802} {\bibfield  {journal} {\bibinfo  {journal} {Phys.
  Rev. A}\ }\textbf {\bibinfo {volume} {69}},\ \bibinfo {pages} {023802}
  (\bibinfo {year} {2004})}\BibitemShut {NoStop}%
\bibitem [{\citenamefont {Machulka}\ \emph {et~al.}(2014)\citenamefont
  {Machulka}, \citenamefont {Haderka}, \citenamefont {Pe\v{r}ina},
  \citenamefont {Lamperti}, \citenamefont {Allevi},\ and\ \citenamefont
  {Bondani}}]{Machulka:14}%
  \BibitemOpen
  \bibfield  {author} {\bibinfo {author} {\bibfnamefont {R.}~\bibnamefont
  {Machulka}}, \bibinfo {author} {\bibfnamefont {O.}~\bibnamefont {Haderka}},
  \bibinfo {author} {\bibfnamefont {J.}~\bibnamefont {Pe\v{r}ina}}, \bibinfo
  {author} {\bibfnamefont {M.}~\bibnamefont {Lamperti}}, \bibinfo {author}
  {\bibfnamefont {A.}~\bibnamefont {Allevi}}, \ and\ \bibinfo {author}
  {\bibfnamefont {M.}~\bibnamefont {Bondani}},\ }\href {\doibase
  10.1364/OE.22.013374} {\bibfield  {journal} {\bibinfo  {journal} {Opt.
  Express}\ }\textbf {\bibinfo {volume} {22}},\ \bibinfo {pages} {13374}
  (\bibinfo {year} {2014})}\BibitemShut {NoStop}%
\bibitem [{\citenamefont {Ostermeyer}\ \emph {et~al.}(2009)\citenamefont
  {Ostermeyer}, \citenamefont {Korn}, \citenamefont {Puhlmann}, \citenamefont
  {Henkel},\ and\ \citenamefont {Eisert}}]{Ostermeyer09}%
  \BibitemOpen
  \bibfield  {author} {\bibinfo {author} {\bibfnamefont {M.}~\bibnamefont
  {Ostermeyer}}, \bibinfo {author} {\bibfnamefont {D.}~\bibnamefont {Korn}},
  \bibinfo {author} {\bibfnamefont {D.}~\bibnamefont {Puhlmann}}, \bibinfo
  {author} {\bibfnamefont {C.}~\bibnamefont {Henkel}}, \ and\ \bibinfo {author}
  {\bibfnamefont {J.}~\bibnamefont {Eisert}},\ }\href {\doibase
  10.1080/09500340903359962} {\bibfield  {journal} {\bibinfo  {journal}
  {Journal of Modern Optics}\ }\textbf {\bibinfo {volume} {56}},\ \bibinfo
  {pages} {1829} (\bibinfo {year} {2009})},\ \Eprint
  {http://arxiv.org/abs/http://dx.doi.org/10.1080/09500340903359962}
  {http://dx.doi.org/10.1080/09500340903359962} \BibitemShut {NoStop}%
\bibitem [{\citenamefont {Yun}\ \emph {et~al.}(2012)\citenamefont {Yun},
  \citenamefont {Xu}, \citenamefont {Zhao}, \citenamefont {Gong}, \citenamefont
  {Bai}, \citenamefont {Shi},\ and\ \citenamefont {Zhu}}]{YunCEPR12}%
  \BibitemOpen
  \bibfield  {author} {\bibinfo {author} {\bibfnamefont {S.}~\bibnamefont
  {Yun}}, \bibinfo {author} {\bibfnamefont {P.}~\bibnamefont {Xu}}, \bibinfo
  {author} {\bibfnamefont {J.~S.}\ \bibnamefont {Zhao}}, \bibinfo {author}
  {\bibfnamefont {Y.~X.}\ \bibnamefont {Gong}}, \bibinfo {author}
  {\bibfnamefont {Y.~F.}\ \bibnamefont {Bai}}, \bibinfo {author} {\bibfnamefont
  {J.}~\bibnamefont {Shi}}, \ and\ \bibinfo {author} {\bibfnamefont {S.~N.}\
  \bibnamefont {Zhu}},\ }\href {\doibase 10.1103/PhysRevA.86.023852} {\bibfield
   {journal} {\bibinfo  {journal} {Phys. Rev. A}\ }\textbf {\bibinfo {volume}
  {86}},\ \bibinfo {pages} {023852} (\bibinfo {year} {2012})}\BibitemShut
  {NoStop}%
\bibitem [{\citenamefont {Fl\'orez}\ \emph {et~al.}(2015)\citenamefont
  {Fl\'orez}, \citenamefont {Calder\'on}, \citenamefont {Valencia},\ and\
  \citenamefont {Osorio}}]{Florez15}%
  \BibitemOpen
  \bibfield  {author} {\bibinfo {author} {\bibfnamefont {J.}~\bibnamefont
  {Fl\'orez}}, \bibinfo {author} {\bibfnamefont {O.}~\bibnamefont
  {Calder\'on}}, \bibinfo {author} {\bibfnamefont {A.}~\bibnamefont
  {Valencia}}, \ and\ \bibinfo {author} {\bibfnamefont {C.~I.}\ \bibnamefont
  {Osorio}},\ }\href {\doibase 10.1103/PhysRevA.91.013819} {\bibfield
  {journal} {\bibinfo  {journal} {Phys. Rev. A}\ }\textbf {\bibinfo {volume}
  {91}},\ \bibinfo {pages} {013819} (\bibinfo {year} {2015})}\BibitemShut
  {NoStop}%
\end{thebibliography}%

\end{document}